A survey is presented of the dynamic features of non-itinerant off-center defects in crystals, such as rotation-like reorientation of isolated species by either impurity or host ions. The occurrence of off-center displacements in electron-vibrational mode systems is regarded as due to the vibronic effect. It is extendable to systems in other fields of physics, e.g. nuclear matter. The extension is found to be effective through re-scaling of energies and separations.

1. Preface

We have lately submitted to the arXiv several pre-prints on the off-center defects in crystals [1-4], both static and itinerant ones, including a short survey on the itinerant species [5]. One should distinguish between the latter species, which we call vibronic or off-center polarons, as opposed to static species, which may comprise both impurity and host ions. The vibronic polaron is a charge carrier coupled to its self-created Jahn-Teller or vibronic distortion. From this point of view, the appearance of both species is obliged to their common generator, which is the (pseudo)-Jahn-Teller effect, that is, the mixing of nearly degenerate electronic states by a coupled phonon or vibrational mode.

We feel that time has come to pay a greater attention to the dynamic features of the static component too, especially in view of our attempts to extend the matter to other systems outside the realm of the electron-phonon multitude. Indeed, we have suggested that at least some quantum-mechanical laws may be common, albeit with certain specific modifications, to condensed matter (soft and hard) and to nuclear matter too, though on a very different energy scale. This extension may hold true to vibronic polarons as well as to their non-itinerant local counterpart which may appear in both condensed and nuclear matter.

The present survey will be expected to complement our earlier work on the static defects which appeared in the literature about a quarter of a century ago [6].

2. Electron-phonon systems

2.1. Off-center effect in dilute systems

Examples for static off-center defects are provided by several smaller-size cation or anion impurities in alkali halides substituting for the respective host ions [6]. For the first group of examples one can place $Li^+$ in KCl and $Cu^+$ in KBr, $Ag^+$ in RbCl, the second group includes $F^-$ in RbBr. Despite its increasing radius, $Ga^+$ goes surprisingly off-center in the excited electronic state of an alkali halide [7]. Another interesting example is the effect of going off-center on $Cu^+$ diffusion [8] and on the absorption band of off-center $Li^+$ in alkali halide [9].

The Hamiltonian of a local two-state electron-phonon (e-ph) system to describe an off-center defect is formally analogical to the band (e-ph) Hamiltonian, only use is made of two local nearly-degenerate electronic states rather than two nearly-degenerate electronic bands [5]:

$$H = H_e + H_L + H_{e\text{-ph}} \qquad (1)$$

where

$$H_e = \Sigma_{m\alpha} \pm \tfrac{1}{2} E_{m\alpha\beta}\, a_{m\alpha}^\dagger a_{m\alpha} \qquad (2)$$

is the electronic counterpart ($\tfrac{1}{2}E_{m\alpha\beta}=|E_{m\alpha}-E_{m\beta}|$: local electron energy gap, $a_m$: electron annihilation operators, $\alpha = \alpha_1,\alpha_2$ labels the electronic states involved), the $m$ summation is over the normal lattice sites active in the process.

$$H_L = \Sigma_m \tfrac{1}{2}[(\mathbf{P}_m^2/M_m) + (M_m\omega_m^2)Q_m^2] \qquad (3)$$

is the harmonic lattice energy diagonalized in the active-mode coordinate $Q_m$. To simplify the matter, we discard possible e-ph couplings that could lead to dependences on the electronic states of the lattice operators. (Such may be found in Reference [10].)

The e-ph interaction ($e_1 \xrightarrow{ph} e_2$ mixing) Hamiltonian is:

$$H_{e\text{-ph}} = \Sigma_{i\alpha}[G_{i\alpha\beta}Q_{i\alpha\beta}+\Sigma_{jk\alpha\beta}D_{ijk\alpha\beta}Q_iQ_jQ_k]\,(|\alpha_1\rangle\langle\alpha_2| + h.c.) \qquad (4)$$

The coupling constants $G_{m\alpha\beta}$ and $D_{ijk\alpha\beta}$ is the heart of the matter. For our previous work we derived simple versions by computing the change of local electrostatic energy as the ion goes off-center: For extending the above reasoning to nuclear systems care will be taken to derive an appropriate substitute. For further applications an isotropic medium will be assumed discarding the site indices $m$ of all parameters.

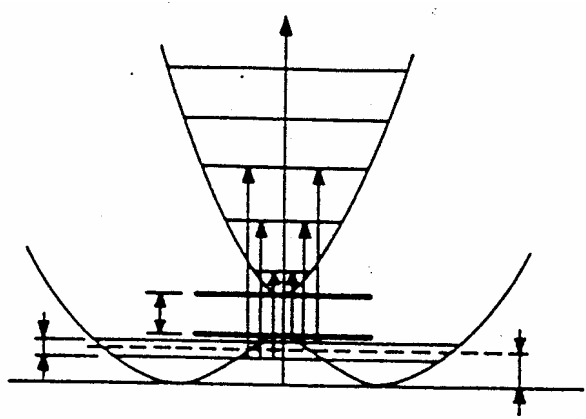

Figure 1. The two-site adiabatic (*alias* vibronic) potential energy surface in the localized limit. The two static electronic levels are shown by thick horizontal segments. These states are mixed by the relevant phonon coordinate along the abscissa to form a local polaron band below the barrier top. Both the polaron mid-band energy and bandwidth are indicated. The vertical arrows suggest optical transitions giving rise to a specific absorption spectrum.

Interwell flip-flop transitions through tunneling or classical jumps redistribute the well population but do not guarantee itinerancy, the latter depending on the well coupling to the nearest-neighboring medium. A multi-well extension is straightforward.

As shown earlier [11], the 1$^{st}$ order mixing constant G drives the ion off-center, the radius of the off-center displacement ring being (K = M$\omega^2$)

$$Q_0 = \sqrt{(4G^4 - E_{\alpha\beta}^2 K^2)}/2GK \qquad (5)$$

obtained from the adiabatic potential energy surface (APES) at all $D_{ijk}$ = 0. For the off-center problem APES of a two-site problem is of the sombrero hat type with the normal lattice site at the central peak and the brim displaced at $Q_0$ where the off-center sites are situated. To linear approximation the brim is a smooth valley where the off-centered particle rotates unimpeded. However, as some 3$^{rd}$ order mixing constants $D_{ijk} \neq 0$ are switched on, the smooth life is over as the brim becomes embroidered by peaks and dips, the latter occurring as metastable wells at the reorientation sites, the former occurring as inter-well barriers in-between..

Once gone off-center, the displaced ion performs reorientation through classic jumps or tunneling transitions across reorientation barriers between off-center sites in which transitions the actor moves upon an off-center surface, e.g. sphere or ellipsoid, or a more complex one [11]. If this rotation-like reorientation occurs within a plane, the planar motion of an off-centered impurity may go along an off-center ring [12]. We have earlier deduced an angular wave equation which controls the reorientation of the displaced impurity:

$$-(\hbar^2/2I)\Delta(\theta,\varphi)\psi \pm (M\omega^2/G)Q_0^4(D_c - D_b)[(\cos\varphi\sin\theta)^4 + (\sin\varphi\sin\theta)^4 + (\cos\theta)^4]\psi - [E - C_\pm]\psi = 0$$

$$C_\pm = D_b + E_{JT}[(1\pm 2) - (E_{\alpha\beta}/4E_{JT})^2] \qquad (6)$$

Here I is the moment of inertia of the rotor, M is the mass and ω is the angular frequency of the coupled oscillator, $Q_0$ is the off-center radius, $D_c < D_b$ are 3$^{rd}$ order e-ph coupling constants, $E_{JT}$ is Jahn-Teller's energy, $E_{\alpha\beta}$ is the energy gap of the electronic states, its bare value short of phonon dressing. Further details can be found in the original publications [13]. No solution to equation (6) has been available so far.

In equatorial plane, setting θ = ½ π and $\Delta(½\pi,\varphi) \equiv \partial^2/\partial\varphi^2$, which may come to be true along the off-center brim of a sombrero-like vibronic potential energy surface, we arrive at Mathieu's equation for the eigenstates of a non-linear oscillator:

$$-(\hbar^2/2I)(\partial^2\psi/\partial\varphi^2) \pm (M\omega^2/b)Q_0^4(D_c - D_b)[(\cos\varphi)^4 + (\sin\varphi)^4]\psi - [E - C_\pm]\psi = 0 \qquad (7)$$

which has to be rearranged to assume its traditional form [11]. This is Schrödinger's angular equation for rotation along the off-center ring of a displaced system. The ring comes as the classical path of an entity rotating around the normal lattice site which entity bears a finite orbital angular momentum. It should be stressed that from a quantum-mechanical viewpoint the off-center rotation is in fact the classical analogue of the splitting of the rotational states of the individual sites due to inter-site tunneling. What we have is the smearing of the off-center entity within an off-center volume as a whole system common for all reorientation sites. The finite angular momentum of the rotating entity would justify the application of chirality as a dynamical characteristics of the off-center defect. On its hand, chirality further points to a helical classical path to connect neighboring off-center planes, a matter already discussed in Reference [5] as regards the vibronic polaron.

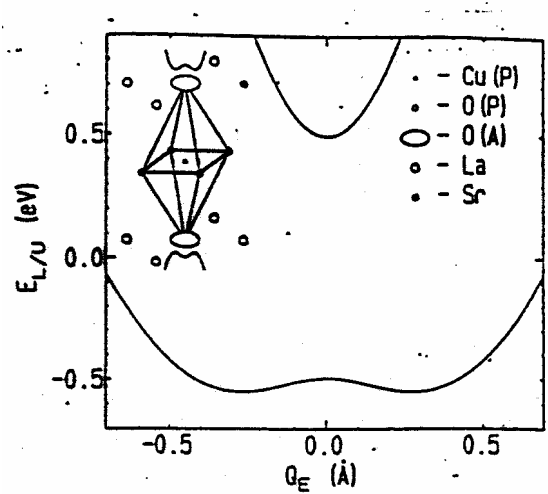

Figure 2. The apical oxygens in the $La_{2-x}Sr_xCuO_4$ high-Tc superconductor believed at a time to play a mediating role for the pairing of in-plane holes, due to its high electric polarizability. This polarizability deforms the shape of the ion turning it oblate. These processes may play a major role for the electric transport along the c-axis (vertical). The particular though typical double-well diagram sets an energy vs. coordinate scale of 1 eV vs. 0.1 Å.

A related observation is that the measured higher $Cu^+$ and $Li^+$ diffusion coefficients in KCl, exceeding the corresponding self-diffusion coefficients by 3-4 orders of magnitude, may be enhanced considerably by dynamic vibronic effects [8]. Indeed, the enhancement is easily understood if the diffusing entity goes off-center. We remind that an off-centered ion is in fact smeared within its own off-center volume, due to quantum-mechanical tunneling between neighboring reorientation sites and the related tunneling level splitting. The smearing increases the effective ion size and its overlap with the diffusion-active sites within a crystal. A controversial hypothesis that $Cu^+$ diffusion proceeds by means of an interstitial mechanism is unlikely in an alkali halide.

One way or the other one, no direct experimental measurements have been performed so far to manifest the *rotation* of off-centered impurity ions in crystals. However, the reorientation *tunneling splitting* has been measured, though not directly [14]. We have already stated that the two qualities are interrelated. Another way of checking the rotational premises is by measuring the orbital magnetic moments [15]. Indeed, it would be fascinating to measure magnetic fields coming from arrangements of off-center dipoles in alkali halides! Appropriate arrangements can be constructed, as mentioned earlier [16], in an $F_A$ center containing KCl crystal where the color centers are all arranged in <110> planes by means of combined photo- and thermo- excitations making use of $F_A$ photo-orientation by $F_{A2}$-polarized light [17]. Once in the <110> planes, the F centers will bring $Li^+$ ions along with, just beneath those planes which puts $Li^+$ arranged too. The result is masterminding an aligned $F_A(Li^+)$ centered crystal to start measurements of possible $Li^+$ related magnetic dipoles.

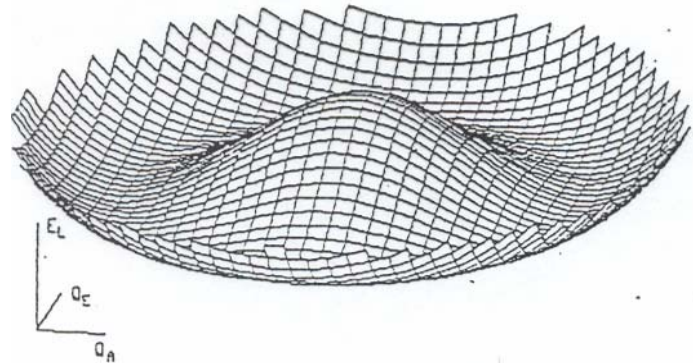

Figure 3. The sombrero-hat vibronic (*alias* adiabatic) potential energy surface of the pseudo-Jahn-Teller mixing. Both the central barrier and the hat brim are seen pointing to a prolate deformation due to $Q_{A0} > Q_{E0}$. The peaks and valleys embroidered along the brim are not shown to avoid overfilling. In a typical case there are four valleys with four interwell barriers in-between. Unless the central barrier is too high, the rotating ion will be smeared within the off-center ellipse.

Other static defect features, such as paraelectric resonance, electron spin resonance and the optical absorption spectra of isolated defect centers have been reviewed comprehensively elsewhere [13].

### 2.2. Concentration dependences in dense off-center systems

A different picture has emerged when studying the dynamic features of denser off-center systems, such as {Li} in KCl. We will now comment on them briefly. In particular, the dielectric-loss relaxation rate has proved concentration-dependent [18]. An example is shown in Figure 4 therein where the reported rates are plotted against the temperatures between 5-200 mK at various Li concentrations and $^6$Li-$^7$Li isotope compositions. These rates may be seen to remain temperature-insensitive below 20 mK roughly followed by a steeper rise with temperature T above 50 mK. This behavior is typical for the relaxation by a rate process which is controlled by quantum-mechanical tunneling at the lowest T, followed by activated tunneling at intermediate T followed by classic jumps at the highest T, as in the reaction rates.

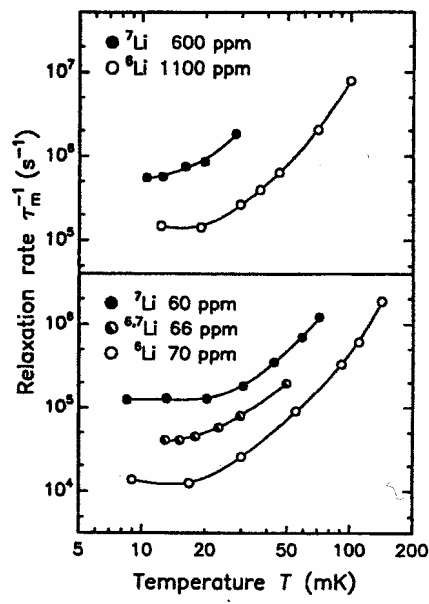

Figure 4: Dielectric loss relaxation rate $\tau_m^{-1}$ vs. T of KCl crystals doped with different Li-isotope admixtures. Solid lines are guides for the eye. After Reference [18].

Improvements to the reaction rate theory [19] have been made lately accounting for both the horizontal tunneling and the vertical-tunneling multi-phonon processes [20]. What is essential now is why the low-temperature rate $\tau_m^{-1}(T)$ is concentration-dependent in that it increases as {Li} is increased. (The Li isotope trend seems another matter.) Nevertheless, the reaction rate theory gives $\tau_{as}^{-1}(T) = \tau_s^{-1}(T) \exp(H/k_B T)$ for the asymmetric relaxation rate of an isolated Li center where H is the zero-point reaction heat. (H > 0 for an exothermic reaction, H < 0 for an endothermic reaction. $\tau_s^{-1}(T)$ is the symmetric rate for an isothermic reaction at H ≡ 0.)

Assuming the Van der Waals dipolar coupling of individual defects, $H \propto \frac{1}{2} t_0 (\alpha/\kappa R_{ij}^3)^2$ where $\alpha$ is the vibronic polarizability of off-center defects, $t_0$ is the ground state tunneling splitting, $\kappa$ is a dielectric constant, $R_{ij}$ is the average inter-center separation at H. We further take for granted that the relaxation rate at 60 ppm is the one characteristic of the residual coupling of isolated centers. Now, the rate at 1100 ppm (hollow circles) is seen to rise tenfold (due to a small increase of H) which is due to a tiny concentration induced decrease of the inter-pair separation as well as a small change of the tunneling splitting $t = t_0(1 + \frac{1}{2} t_E/t_0)$ where $t_E$ is the field-induced level splitting.

### 3. Fermion-boson systems

For calculated e-ph coupling constants of the order of 1 eV/Å effective at distances as large as 0.1 Å, the expectation for a nuclear fermion-boson (f-b) system will be of the order of 1 MeV/Fermi where 1 Fermi = $10^{-13}$ cm. This will hold good as the harmonic oscillator method is applicable to nuclear matter, as assumed [21]. The transfer of solid state language to nuclear matter is not new and has already proved feasible since Yukawa times. Then the Coulomb interaction of two charged bodies via the exchange of photons has been transferred to the nuclear forces between neutrons and protons assuming the exchange of heavier quanta, now identified as π-mesons [22].

### 3.1. First-order f-b coupling constant

For calculating a linear f-b coupling constant G, the quantum-mechanical Yukawa potential of a short-range force may be of use. We postulate $\phi(r) = \exp(-\kappa r)/r$ with $\kappa = 5 \times 10^{12}$ cm$^{-1}$ [23]. We also incorporate a boson $A_{1g}$ mode coordinate Q by way of

$\phi(r+Q) = \exp(-\kappa[r+Q])/(r+Q)$.

Other choices may be found closer to real systems. Differentiating in Q we get

$\partial\phi/\partial Q = \{\exp(-\kappa[r+Q]) / (r+Q)^2\} \{-\kappa(r+Q)-1\}$

which yields

$G(r) \equiv \partial\phi/\partial Q |_{Q=0} = -(1+\kappa r) \exp(-\kappa r) / r^2$  (8)

This is the linear f-b operator. The coupling constants obtain in the form of diagonal or off-diagonal matrix elements $G_{\alpha\beta} = \langle\alpha|G(r)|\beta\rangle$. We are now interested in the mixing constant at $\alpha \neq \beta$. The fermion states to be mixed by means of $G_{\alpha\beta}$ need not be eigenstates of the Yukawa potential. However, if they ought to be, there is a source to providing solutions by analogy with the Debye-Hückel screening in electrostatics, as described by a similar equation [24].

One way or the other, we may attempt to use spherical-well Bessel eigenfunctions for obtaining order of magnitude estimates [25]:

$J_0(\kappa r) = A_{\kappa 0}\sin(\kappa r)/\kappa r$
$J_1(\kappa r) = A_{\kappa 1}(\kappa r)^{-1}[\sin(\kappa r)/(\kappa r) - \cos(\kappa r)]$, etc.  (9)

where $A_{\kappa 0}$ and $A_{\kappa 1}$, etc. are normalization constants.

Using spherical Bessel functions, finite diagonal matrix elements have been computed [10], though with a different linear-coupling operator, as shown below. The diagonal matrix elements produce (a)diabatic electronic energies. However, it may be seen at first glance that the linear-coupling operator (8) would produce a divergent 1st-order constant. For an estimate of the off-diagonal (mixing) matrix element, we set $\alpha = 0$, $\beta = 1$:

$G_{01} = G_{10} = \langle j_0(\kappa r)|G(r)|j_1(\kappa r)\rangle = -\kappa \int_0^\infty du\, [\sin u / u - \cos u] \sin u\, (1/u)^2 (1+u) \exp(-u)/u^2$

$= -\kappa \int_0^\infty du\, \{(1/u)^5 \sin u + (1/u)^4 [\sin u - \cos u] - (1/u)^3 \cos u\} \sin u \exp(-u)$  (10)

which diverges, due to the singularity at the center of the well (r = 0). This shows that spherical Bessel functions may not be mixed by the $A_{1g}$ phonon mode operator (8) across the Yukawa potential.

On the other hand, the spherical-well potential is defined by

$V(r) = V_0$ at $0 \leq r \leq r_0$, $V(r) = 0$ at $r > r_0$.  (11)

Its eigenstates are exemplified by the lowest two spherical Bessel functions (9). The Q-derivative of the spherical potential coupled to the $A_{1g}$ breathing mode coordinate is

$dG(r)/dQ|_{Q=0} = -\Delta\delta(r-r_0)$  (12)

The corresponding coupling constants are

$G_{00} = -\Delta A_{\kappa 0}^2[\sin(\kappa r_0)/(\kappa r_0)]^2$

$G_{11} = -\Delta A_{\kappa 1}^2\{(\kappa r_0)^{-1}[\sin(\kappa r_0)/(\kappa r_0) - \cos(\kappa r_0)]\}^2$

$$G_{12} = G_{21} = \Delta A_{\kappa 0} A_{\kappa 1}[\sin(\kappa r_0)/(\kappa r_0)]\{(\kappa r_0)^{-1}[\sin(\kappa r_0)/(\kappa r_0) - \cos(\kappa r_0)]\} \quad (13)$$

where we set $\kappa = 5\times 10^{12}$ cm$^{-1}$, while $A_{\kappa 0}$ and $A_{\kappa 1}$ are normalization constants. We perform the calculations devised by (13) at $r_0 = 10^{-13}$ cm = 1 Fermi to obtain numerical values for G. The other kinetic parameters ought to be known for constructing a theory are $K_n = M_n \omega_n^2$ in which $M_n$ is the nucleon mass and $\omega_n$ is its frequency. If known, these parameters may be utilized for finding the adiabatic potentials at (00) and (11) (cf. (13)). The diabatic parabola is (in 1-D):

$$V_{00}(Q) = \tfrac{1}{2} KQ^2 - G_{00}Q + E_{00}$$

$$V_{00}(Q) = \tfrac{1}{2} KQ^2 - G_{11}Q + E_{11} \quad (14)$$

while the mixing 1-D double (two-site) well reads

$$V_{12}(Q) = \tfrac{1}{2} KQ^2 \pm \tfrac{1}{2} \sqrt{[(2G_{12}Q + D)^2 + E_{gap}^2]} \quad (15)$$

Here $E_{gap} = |E_{00} - E_{11}|$ is the gap energy, D is an asymmetry parameter amounting to a constant Q-independent mixing term. D = 0 for symmetric wells, as in rotational problems (cf. Figure 1). The next step is calculating Jahn-Teller's energy $E_{JT} = G_{12}^2/2K$ and the off-center radius $Q_0 = \sqrt{(4G_{12}^4 - K^2 E_{gap}^2)}/2G_{12}K = \sqrt{(2E_{JT}/K)}\sqrt{[1 - (E_{gap}/4E_{JT})^2]}$. The multi-site rotational surface $V_{ij}(Q)$ obtains by generalizing the two-site one (15) and is composed of a multitude of peaks and valleys, not appearing in Figure 3, which are situated along the sombrero-hat brim. The peak- and valley- embroidered brim is 3$^{rd}$-order in the mixing constants [26]. The exact form of the multi-valley surface can be found elsewhere [2]. The valleys are metastable wells which correspond to certain important orientation sites in the active medium.

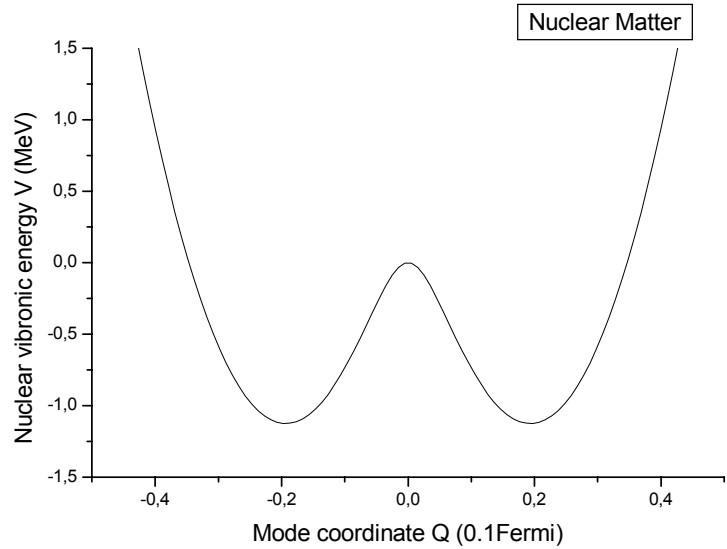

Figure 5. As suggested by Figure 2, the symmetric double-well diagram appropriate for a nuclear vibronic process will set a perspective energy vs. coordinate scale of 1 MeV vs. 0.1 Fermi. We assume G = 10 MeV/F, K = 100 MeV/F$^2$ to get $E_{JT}$ = 0.5 MeV and select a gap energy of $E_{gap}$ = 1 MeV to draw the strong-coupling ($4E_{JT} > E_{gap}$) diagram of the Figure 1

type (lower branch), as above. This might be the two-site vibronic energy of a proton in nuclear matter, represented as a system of harmonic oscillators. The diagram shows the central peak configuration at $Q = 0$ unstable, while two metastable off-center configurations at $Q = \pm 0.2$ Fermi appear to set a 1-D analogue of the off-centered sombrero brim in Figure 3.

The form of sombrero brim depends on the symmetry of the mixing mode. It is a circular ring for symmetric modes such as $T_{1u}$. Accordingly, the off-centered surface will be a sphere, though it may turn oblate or prolate for lower mode symmetries. This provides a simple explanation for the observed deformed nuclei, as it certainly does for solid state ions [27]. Namely the apparent nucleus form depends on the shape of off-center volume it is smeared within. The boson coordinate is defined by $Q_n \propto (b_n^\dagger + b_n)$ where the $b_n$'s are the boson ladder operators. An example for a deformed oblate shape may be identified in Figure 2 where apical oxygen ions in $La_{2-x}Sr_xCuO_4$ presumably serve as polarizable partners in the pairing of in-plane holes [28].

For comparison, the perspective adiabatic potential energy surface of nuclei is shown in Figure 5 as deduced from the above data on the mixing constant $G = 10$ MeV/F and from setting $K = M_N\omega_N^2 = 100$ MeV/F$^2$ to get $E_{JT} = 0.5$ MeV and select a gap energy of $E_{gap} = 1$ MeV to draw the strong-coupling ($4E_{JT} > E_{gap}$) diagram, from these acceptable values [29]. In fact, this is rescaling which indicates that the physical law is the same whether in solid state or in nuclear matter: we arrive at a remarkable conclusion!

Finally, we would like to comment on the supersymmetry and its violation through coupling to Higgs bosons [30]. The ideal supersymmetric configuration is the one in which each half-integer spin has a partner of full-integer spin. The Higgs boson being one such partner of spin 0, it lowers the symmetry at the fermion site through a Jahn-Teller type distortion. This type of symmetry breaking could occur in an atomic cluster or even in the gaseous phase, or in soft condensed matter too. What remains to be seen in LHC experiments is to find out at what stage does a symmetry breaking occur.

4. Conclusion

The focal point to argue on this survey is the extension of solid state problems to nuclear matter related applications. The interdisciplinary studies have often been fruitful, as with Yukawa's forces, to mention a few. This time we extended the quantum-mechanical off-center effects, first established for solid state systems doped with smaller size impurity ions, and then even to host ions in high-Tc superconductors. We believe that identifying the electron and phonon partners as fermion-boson pairs gives an unique opportunity to widen the realm of solid state borrowed items to nuclear physics and perhaps to optics too. The off-centered extension makes it possible to explain in simple terms the deformed nuclei (oblate, prolate, and cigar-shaped) as reflecting the symmetry of the fermion-coupled boson mode.

The vibronic or off-center polarons in e-ph solid problems provide another interdisciplinary example in that they are directly extendable to f-b coupling in nuclear matter. We remind of the boson dressing of fermions which has long been introduced in nuclear problems, though it may again have been deeply rooted in the solid state test ground.

Another key problem is that of the supersymmetry breaking which could again be tackled by solid state means and, indeed, one questions the restricted applicability of Jahn-Teller or vibronic effects to condensed matter alone and not to a wider field of science. If so, the scientific implications could be miraculous.

Half a century before, solid state models have been regarded as a simpler test ground for verifying new theories and developing new techniques to ultimately solving nuclear problems. Now the trend may be reversed in that solid state methods suggest solutions to enigmatic features of nuclear physics.

References

[1] G. Baldacchini, R.M. Montereali, U.M. Grassano, A. Scacco, P. Petrova, M. Mladenova, M. Ivanovich, and M. Georgiev, cond-mat 0709.1951.
[2] G. Baldacchini, R.M. Montereali, U.M. Grassano, A. Scacco, P. Petrova, M. Mladenova, M. Ivanovich, and M. Georgiev, physics 0710.0650.
[3] G. Baldacchini, R.M. Montereali, U.M. Grassano, A. Scacco, P. Petrova, M. Mladenova, M. Ivanovich, and M. Georgiev, physics 0710.2334.
[4] G. Baldacchini, R.M. Montereali, U.M. Grassano, A. Scacco, P. Petrova, M. Mladenova, M. Ivanovich, and M. Georgiev, physics 0710.3583.
[5] Mladen Georgiev, cond-mat arXiv: 0809.5285.
[6] A. Diaz-Gongora, C. Medrano P., J.L. Boldu O., R.J. Gleason, and M. Georgiev, Revista Mexicana de Fisica **32** (3) 475-504 (1986).
[7] K. Kalder, V. Korrovits, V. Nagirnyi, Z. Zazubovich, Phys. Status Solidi B **178**, 391 (1993); V. Nagirnyi, A. Stolovich, Z. Zazubovich, N, Janson, Phys. Status Solidi B **173**, 743 (1992).
[8] F. Despa, ICTP Preprint 10/94/176, Trieste, Italy.
[9] M. Leblans, Thesis, University of Antwerp (U.I.A), Antwerpen, Belgium (1990).
[10] M. Georgiev, J. Inform. Record. Mater. **13** (2) 75-93 (1985); **13** (3) 177-195 (1985); **13** (4) 245-256 (1985).
[11] P.C. Petrova, M.D. Ivanovich, M. Georgiev, M.S. Mladenova, G. Baldacchini, R.-M. Monterali, U.M. Grassano, and A. Scacco, Proc.13th Internat. Conference on Defects in Insulating Materials: ICDIM '96, G.E. Matthews and R.T. Williams, eds., Winston-Salem NC, 1996 (Materials Science Forum, Volumes 239-241, 1997) p.p. 377-380:
[12] F. Lüty, J, Physique (Supplement) **34** (11-12) C9-49-59 (1973).
[13] M.D. Glinchuk in: *The dynamical Jahn-Teller effect in localized systems*, Yu.E. Perlin and M. Wagner, eds. (Elsevier, Amsterdam, 1984), pp. 819-872.
[14] V. Narayanamurti and R.O. Pohl, Revs. Mod. Phys. **42**, 201 (1970).
[15] A.G. Andreev, M. Georgiev, M.S. Mladenova, and V. Krastev, Internat. J. Quantum Chem. **89**, 371-376 (2002).
[16] F. Luty in: *Physics of color centers*, W.B. Fowler, ed. (Academic, New York, 1968):
[17] M. Georgiev, M. Mladenova, V. Krastev, and A. Andreev, Eur. Phys. J. B **29**, 273-277 (2002).
[18] C. Enss, M. Gaukler, M. Nullmeier, R. Weis, A. Würger, Phys. Rev. Lett.**78**, 370 (1997)
[19] S.G. Christov, Phys. Rev. B **26,** 6918 (1982).
[20] M. Georgiev and A. Gochev, Internet Electronic J. of Molecular Design, **4** (12) 862-881 (2005).
[21] D.J. Rowe, *Nuclear collective motion* (Methuen, London, 1970).
[22] H.A. Bethe, F. Hoffmann, *Mesons and Fields*, vol. II *Mesons* (Row, Peterson & Co.,

New York, 1955).
[23] K, Nishijima, *Fundamental particles* (Benjamin, Amsterdam, 1964)
[24] V.L. Bonch-Bruevich and V.B. Glasko, Optika i Spectroskopiya **14** (4) 495 (1963).
[25] L.I. Schiff, *Quantum Mechanics* (McGraw Hill, Tokyo, 1987).
[26] M.D. Glinchuk, M.F. Deigen, A.A. Karmazin, Fiz. Tverdogo Tela **15**, 2048 (1973).
[27] Mladen Georgiev, physics arXiv:0802.2538.
[28] M. Georgiev, Revs. Solid State Sci. **5** (4) 551-567 (1991).
[29] A. Bohr, B.R. Mottelson, *Nuclear structure* vol. I *Single particle motion* (Benjamin, New York, 1969).
[30] An easy source: *The Higgs boson* from Wikipedia, Internet 2008.